\documentclass[11pt]{article}
\usepackage[margin=1in]{geometry}
\usepackage{natbib}
\usepackage{amssymb}
\providecommand{\email}[1]{\thanks{\texttt{#1}}}
\providecommand{\pagerange}[1]{}
\providecommand{\pubyear}[1]{}
\providecommand{\volume}[1]{}
\providecommand{\artmonth}[1]{}
\providecommand{\doi}[1]{}
\providecommand{\backmatter}{}
\newenvironment{keywords}{\par\medskip\noindent\textbf{Key words:} }{\par\medskip}
\date{}

\usepackage[figuresright]{rotating}
\usepackage{tikz}
\usepackage{amsmath}
\usepackage[hyphens]{url} 
\usepackage{hyperref}
\usepackage[utf8]{inputenc}
\usepackage{graphicx}
\usepackage{longtable}
\usepackage{booktabs}

\title{Cross-Fitted
Survey-Weighted TMLE with Design-Based Variance for Causal Machine
Learning}

\author{ M. Ehsan
Karim \email{\href{mailto:ehsan.karim@ubc.ca}{\nolinkurl{ehsan.karim@ubc.ca}}} \\ School
of Population and Public Health, University of British Columbia,
Vancouver, British Columbia, Canada, and Centre for Advancing Health
Outcomes, St.~Paul's Hospital, Vancouver, British Columbia, Canada 
	   }

\usepackage{array}
\usepackage{here}
\usepackage{lscape}
\author{M. Ehsan Karim\\[4pt]
\normalsize
\begin{minipage}{0.82\textwidth}\centering
School of Population and Public Health, University of British Columbia, Vancouver, British Columbia, Canada; and Centre for Advancing Health Outcomes, St.\ Paul's Hospital, Vancouver, British Columbia, Canada\\[3pt]
\texttt{ehsan.karim@ubc.ca}
\end{minipage}}

\begin{document}

\date{}

\pagerange{\pageref{firstpage}--\pageref{lastpage}} \pubyear{}

\volume{0}
\artmonth{January}
\doi{0000-0000-0000}


\label{firstpage}


\maketitle

\begin{abstract}
Cross-fitting is not a refinement of survey-weighted causal machine
learning but, once the nuisances are flexible, what restores valid
inference. We study the population average treatment effect under a
stratified multistage design, estimated by a survey-aware targeted
maximum likelihood estimator (TMLE) whose variance is obtained by
Taylor-series linearization of the influence function, treating the
primary sampling unit as the replication unit. Our central result is
that this validity turns on cross-fitting at the cluster level:
sufficiency is established in theory, and the failure without it is
shown in simulation. Once flexible learners cross a complexity (Donsker)
boundary, single-fit survey TMLE can severely under-cover, and internal
cluster-aware cross-validation does not substitute for cross-fitting;
among the estimators we evaluate, only out-of-fold fitting at the
cluster level restores valid coverage. In simulations spanning a
many-PSU and an NHANES-like design, on a diverse ensemble the single-fit
and internal cross-validation estimators cover at about 0.89--0.91 and
0.85--0.88 while the cross-fitted estimator holds at 0.93--0.95, and an
aggressively grown learner drives single-fit coverage to 0.18--0.22. Two
scope choices are deliberate: survey-weighted point estimation is prior
work, and the nuisance product-rate condition is assumed and probed
empirically. Within these conditions we prove asymptotic normality and
design-consistency of the linearization variance. Four NHANES analyses
and open-source software illustrate the method.
\end{abstract}

%
%


\section{Introduction}\label{introduction}

\textbf{Background and Motivation}: Many large-scale national health
surveys across the globe (e.g., DHS, CCHS, BRFSS) employ complex
sampling designs to ensure population representativeness and analytic
efficiency \citep{heeringa2017applied}. Such designs commonly involve
stratification, multistage clustering, and unequal probabilities of
selection.

In the United States, the National Health and Nutrition Examination
Survey (NHANES) is a leading example of a complex, multistage survey and
is extensively used in epidemiology, clinical research, and public
health. However, the validity of inferences drawn from such analyses
depends critically on accounting for NHANES's sampling design. Standard
statistical methods that assume simple random sampling may yield biased
point estimates and invalid confidence intervals when applied without
proper adjustment for design features such as stratification,
clustering, and weighting.

\textbf{Challenges in Survey-Based Causal Inference}: Targeted Maximum
Likelihood Estimation (TMLE) is a doubly-robust framework for causal
inference that integrates flexible machine learning with semiparametric
efficiency theory \citep{van2006targeted}. However, TMLE was originally
developed under the assumption of independent and identically
distributed (i.i.d.) observations, and its standard implementation does
not accommodate the complexities of complex survey designs.

Adapting causal inference to complex surveys is an active area. Most
progress has been on propensity-score methods
\citep{austin2018propensity}, and a unifying perspective clarifies when
the sample-selection and exposure-selection mechanisms require the
weights to enter each model \citep{nattino2025causal}; efficient
propensity-score-based estimators for population inference from
epidemiologic cohorts have also been developed
\citep{wang2022efficient}. The targeted-learning literature provides the
doubly-robust machinery we build on \citep{van2011targeted}, and TMLE
has been adapted to specific dependence structures such as clustering in
randomized trials and interference
\citep{balzer2019new, nugent2025causal, balzer2023two}. What this body
of work has largely settled is how survey weights enter \emph{point}
estimation; what remains comparatively underdeveloped is valid
\emph{variance} estimation for a doubly-robust, machine-learning
estimator under stratified multistage clustering. In practice, current
software can incorporate observation weights in the point estimate but
does not natively deliver Taylor-linearized standard errors (SEs)
accounting for stratification and clustering \citep{gruber2012tmle}, and
the behavior of flexible learners in this setting without cross-fitting
has not been characterized. The closest existing variance results do not
close it: neither \citet{wang2022efficient} nor
\citet{nattino2025causal} supplies what Theorems 1--2 (Section 4)
deliver---a design-based variance, covered without a Donsker condition,
for a doubly robust estimator with flexible machine-learning nuisances
under stratified multistage clustering---the former giving asymptotic
variances for parametric propensity-score weighting and matching
estimators (no machine-learning nuisances, cross-fitting, or doubly
robust targeting), the latter settling only where the weights enter
\emph{point} estimation (Web Appendix \S A contrasts both in detail).
Three concurrent 2026 works adapt cross-fitting to dependent or survey
data but leave the design-based clustered-causal variance open. The most
relevant, \citet{balkus2026cross}, shows that ``as-independent''
cross-fitting still removes the key bias terms under correlation, so
correlation-aware fold schemes are typically unnecessary---a
point-estimation \emph{negligibility} result that is orthogonal to, not
in tension with, our design-based \emph{necessity} claim, since it
carries no survey weights or stratified multistage design and addresses
neither variance nor confidence-interval construction. The other two
\citep{dagdoug2026machine, qian2026changepoint} adapt cross-fitting to
finite-population model-assisted estimation and to changepoint location,
respectively, reaching neither a causal influence function nor a
PSU-level clustered variance (full contrast in Web Appendix \S A).

\textbf{Contributions}: A valid survey TMLE must address two failures at
once---the design variance induced by stratification and clustering,
which narrows confidence intervals even when the point estimate is
unbiased, and an empirical-process term left by flexible
machine-learning nuisances fit without cross-fitting---and we make three
contributions to that end. First, we show that with flexible
machine-learning nuisances, cross-fitting at the primary sampling unit
(PSU) level is what keeps the variance of a survey-weighted causal
estimator valid: the theory establishes sufficiency---cross-fitting,
plus a finite-population stability condition, restores valid inference
where the single fit is not covered---and simulations spanning a
many-PSU and an NHANES-like design demonstrate the failure without it.
Single-fit survey TMLE can severely under-cover once the learners cross
a complexity (Donsker) boundary, and internal cluster-aware
cross-validation does not substitute for cross-fitting. This failure is
invisible at the parametric tier on which most applied survey-causal
guidance rests, surfacing only once genuinely flexible learners are
used. Second, we develop the survey-weighted TMLE that delivers this and
show, treating the PSU as the independent replication unit, that it is
asymptotically normal with a design-consistent
Taylor-series-linearization variance under stratified multistage
sampling---a result proved under explicit conditions of which the
central one, a product accuracy rate on the machine-learning nuisances,
is an assumption we probe empirically (Section 4). Third, we quantify
the separate consequences of ignoring the weights, which biases the
point estimate under informative sampling, and of ignoring the clustered
variance, which understates uncertainty; we illustrate all three points
on four contemporary NHANES analyses in which respecting the design
changes what the analyst would report---most sharply by removing, at the
survey-weighting step, an apparently significant
e-cigarette--hypertension association. Our contribution is thus the
machine-learning-nuisance, cross-fitted, design-aware-variance extension
of this established survey-causal machinery
\citep{van2011targeted, nattino2025causal, wang2022efficient}, with a
computationally inexpensive linearization SE---not a first treatment of
causal inference in complex surveys.

\textbf{Aims and roadmap}: We extend TMLE variance estimation to survey
settings by integrating semiparametric theory with survey statistics
theory
\citep{van2011foundations, woodruff1971simple, fuller1975regression},
using off-the-shelf open-source software, and illustrate it on a
simulation that reproduces NHANES-like stratification, clustering, and
informative selection and on four contemporary NHANES questions
(2007--2018) spanning exposure prevalence and overlap (Section 6).

\section{Notation, Estimand, and
Assumptions}\label{notation-estimand-and-assumptions}

\textbf{Notation.} The observed unit is \(O=(C,A,Y)\sim P_0\), where
\(C\) is a vector of confounders, \(A\in\{0,1\}\) a binary treatment,
and \(Y\in\{0,1\}\) a binary outcome; write
\(Q_0(a,c)=E_{P_0}(Y\mid A=a,C=c)\) and
\(g_0(1\mid c)=\Pr_{P_0}(A=1\mid C=c)\) for the outcome and propensity
nuisance functions. The survey delivers, for each sampled person \(i\),
the tuple \((O_i,w_i,h_i,j_i)\), where \(w_i=1/\pi_i\) is the known
design weight (inverse inclusion probability), \(h_i\) indexes one of
\(H\) strata, and \(j_i\) indexes a primary sampling unit (PSU) within a
stratum; stratum \(h\) contributes \(m_h\ge 2\) sampled PSUs. We write
\(P_n^w f=\widehat N^{-1}\sum_{i\in S} w_i\, f(O_i)\) for the Hájek
(survey-weighted) average operator, with
\(\widehat N=\sum_{i\in S} w_i\), and \(P_0 f=E_{P_0}\{f(O)\}\) for the
population mean.

\textbf{Estimand.} The target is the population average treatment effect
(ATE) on the risk-difference scale, \(\Psi=E_{P_0}\{Y(1)-Y(0)\}\), where
\(Y(a)\) denotes the potential outcome under treatment level \(a\). We
take \(\Psi\) to be the super-population ATE; the finite-population ATE
of the realized population, \(\Psi_N\), is a secondary target that the
same estimator consistently estimates with the same leading variance,
and the same machinery extends to the risk-ratio and odds-ratio scales
and to bounded continuous outcomes, with a strengthening of the
design-nondegeneracy condition and, for continuous outcomes, a
non-separation condition on each arm's outcome law (Web Appendix \S A,
Corollary 1, and \S D).

\textbf{Causal assumptions.} Identification of \(\Psi\) rests on the
standard conditions (i)--(iii): (i) \emph{consistency}, \(Y=Y(A)\), with
no interference and a single version of treatment (SUTVA); (ii)
\emph{conditional exchangeability},
\(\{Y(0),Y(1)\}\perp\!\!\!\perp A\mid C\); and (iii) \emph{treatment
positivity}, \(\delta\le g_0(1\mid C)\le 1-\delta\) almost surely for
some \(\delta\in(0,1/2)\). We additionally assume (iv) \emph{outcome
overlap}, \(\delta'\le Q_0(a,C)\le 1-\delta'\) almost surely for
\(a\in\{0,1\}\), an estimation-side condition which ensures that
bounding the fitted outcome model away from zero and one does not
distort the target.

\textbf{Sampling assumptions.} Two further conditions connect the sample
to the population. First, \emph{non-informative sampling given the
confounders}: writing \(S\) for the selection indicator,
\(S\perp\!\!\!\perp(A,Y)\mid C\), so that conditional on \(C\) selection
carries no additional information about treatment or outcome
\citep{pfeffermann1993role}. Under this condition the conditional laws
of \(Y\mid A,C\) and \(A\mid C\) are the same in the sample as in the
population, and survey-weighted fits recover the super-population
nuisances \((Q_0,g_0)\), consistent with the survey-causal framework of
\citet{nattino2025causal}. Second, \emph{sampling positivity with a
known design}: inclusion probabilities are known and positive, with the
weights bounded by a fixed multiple of the population-average weight
(bounded relative weights) and the number of sampled individuals per PSU
bounded; the meta-justifications for these forms --- including why a
uniform lower bound on the inclusion probabilities is deliberately not
assumed --- are detailed in Web Appendix \S A. The causal assumptions
(i)--(iii) alone yield the identification result
\(\Psi=E_{P_0}\{Q_0(1,C)-Q_0(0,C)\}\); the two sampling conditions, with
the bounded relative weights, then make this quantity estimable from the
weighted sample (Web Appendix \S A, Proposition 1).

\textbf{Inferential framework.} We adopt the combined model--design
(super-population) framework of \citet{rubinbleuer2005two}, which
reconciles treating the outcome as random, as causal inference requires,
with treating it as fixed, as design-based survey inference does. The
PSU---the ultimate cluster, an independent draw from a cluster-level
super-population within which individuals may be arbitrarily
dependent---is thus the independent replication unit that carries the
asymptotics, and it justifies both the estimand above and the
design-based variance of the next section. The two-layer construction is
treated in full in Web Appendix \S A.

\section{Methods}\label{methods}

\subsection{Survey-weighted TMLE}\label{survey-weighted-tmle}

TMLE estimates the ATE by combining an outcome model with a propensity
model in a doubly robust, semiparametric update \citep{van2006targeted};
we make it design-aware by weighting every stage with \(w_i\); in the
single-fit estimator we also estimate the nuisances by survey-weighted
Super Learner, so that each model's cross-validated risk targets the
population rather than the sample \citep{van2011foundations}, whereas
the cross-fitted estimator instead fits its out-of-fold nuisances
\emph{unweighted} (cross-fitting subsection below). Throughout, \(Y\) is
the binary outcome, \(A\) the binary treatment, \(C\) the confounders,
and \(w_i=1/\pi_i\) the design weight.

\textbf{Nuisance models and targeting.} We fit the outcome model
\(Q_w(A,C)=E[Y\mid A,C]\) and the propensity model
\(g_w(1\mid C)=P(A=1\mid C)\) by maximizing the survey-weighted
log-likelihoods, then update \(Q_w\) to a targeted \(Q_w^*\) by the
standard TMLE fluctuation: a survey-weighted logistic regression of
\(Y\) on the two arm-specific clever covariates \(H_1=A/g_w(1\mid C)\)
and \(H_0=(1-A)/\{1-g_w(1\mid C)\}\) with offset \(\text{logit}\,Q_w\),
whose fitted pair of fluctuation parameters \((\epsilon_0,\epsilon_1)\)
solves the survey-weighted influence-function estimating equation. This
arm-specific parametrization is the one the \texttt{tmle} package fits
and the one our theory covers; the full targeting equations are standard
and are given in Web Appendix \S B. Their difference \(H=H_1-H_0\) is
the influence-function covariate of Equation (\ref{eq:eif}).

\textbf{Point estimate and influence function.} The ATE estimate is the
Hájek-weighted contrast of the targeted predictions (Equation
\ref{eq:ate}), and each unit's influence function is Equation
(\ref{eq:eif}); we say \emph{influence function} rather than
\emph{efficient influence function} throughout; Section 4 explains why.

\begin{equation}
\widehat\psi_w = \frac{\sum_{i=1}^{n} w_i Q_w^*(1, C_i)}{\sum_{i=1}^{n} w_i} - \frac{\sum_{i=1}^{n} w_i Q_w^*(0, C_i)}{\sum_{i=1}^{n} w_i}
\label{eq:ate}
\end{equation} \begin{equation}
D_i = H(A_i, C_i) \cdot (Y_i - Q_w^*(A_i, C_i)) + Q_w^*(1, C_i) - Q_w^*(0, C_i) - \widehat\psi_w
\label{eq:eif}
\end{equation}

\subsection{The design-based variance}\label{the-design-based-variance}

Treating the PSU as the independent replication unit, we estimate the
variance of \(\widehat\psi_w\) by Taylor-series linearization of the
survey-weighted mean of the per-unit influence functions
\citep{woodruff1971simple, fuller1975regression, binder1983-variances},
as in Equation (\ref{eq:var}), where \(\hat{N} = \sum_{i \in S} w_i\),
\(m_h\) is the number of sampled PSUs in stratum \(h\), and
\(\hat{u}_{hj} = \sum_{i \in \text{PSU}(h,j)} w_i(D_i - \overline{D}_{svy})\)
is the within-PSU weighted total of the centered influence function
(stratum-\(h\) average \(\hat{\bar{u}}_h\)). This is the standard
ultimate-cluster variance applied to the influence function, accounting
for weights, stratification, and clustering at once.

\begin{align}
\text{Var}(\widehat\psi_w) &\approx \frac{1}{\hat{N}^2} \sum_{h=1}^{H} \frac{m_h}{m_h - 1} \sum_{j=1}^{m_h} (\hat{u}_{hj} - \hat{\bar{u}}_h)^2 \label{eq:var} \\
\widehat{\text{se}}_w &= \sqrt{\text{Var}(\widehat\psi_w)}\nonumber
\end{align}

A formal proof that \(\widehat\psi_w\) is asymptotically normal and that
this linearization variance is design-consistent, under explicit
conditions of which the central nuisance product rate is a stated
assumption (Section 4), is given in Web Appendix \S A (Theorems 1 and
2). As shown there, the single-fit estimator is covered by these
theorems only in the parametric, fixed-dimensional case; for
richer---even Donsker---nuisance classes we do not establish it (see
``Scope'\,' in Section 4), and flexible learners can leave the Donsker
regime altogether; the cross-fitted estimator of the next subsection is
covered with no such complexity restriction. Web Appendix \S B gives the
R implementation.

\subsection{Cross-fitting
(Fully-Aware-CF)}\label{cross-fitting-fully-aware-cf}

With aggressive Super Learner libraries the single-fit estimator can
fail, because reusing the same observations to fit the nuisances and to
evaluate the influence function leaves an empirical-process term that
need not vanish (Section 4 makes this precise). The cross-fitted
estimator, which we call Fully-Aware-CF, removes this requirement.

The estimator partitions the sampled PSUs into folds at the cluster
level---whole PSUs are assigned to \(V\ge2\) folds, balanced within each
stratum (with \(m_h=2\) PSUs, as in most NHANES strata, this is
leave-one-PSU-per-stratum cross-fitting)---so that each held-out PSU's
nuisances are independent of its own data. For each fold the outcome and
propensity models are fit on the PSUs \emph{outside} the fold by Super
Learner \emph{without} the survey weights and predicted on the held-out
PSUs; under non-informative sampling given the confounders
(\(S\perp\!\!\!\perp(A,Y)\mid C\)) these unweighted out-of-fold fits
target the same \((Q_0,g_0)\) as weighted fits (Web Appendix \S A,
Proposition 1); their consistency and rate remain assumptions on the
learner ((C1), Section 4). The survey weights re-enter at a single
pooled, survey-weighted targeting step---one pair of arm-specific
fluctuations \((\epsilon_0,\epsilon_1)\) fit on all folds---and at the
variance, which is Equation (\ref{eq:var}) evaluated on the resulting
cross-fitted influence function. Because each held-out PSU's offset and
propensity are trained on the other PSUs and only the two pooled
fluctuation parameters see the held-out fold, the empirical-process term
\(T_1\) is controlled with no Donsker condition on the learners (Web
Appendix \S A). The full recipe, the truncation constants, and the fold
construction are in Web Appendix \S B.

\section{Asymptotic Theory}\label{asymptotic-theory}

\textbf{The primary sampling unit as the unit of inference.} Asymptotics
are taken in the number of sampled PSUs \(m=\sum_h m_h\), the
independent units; the two design regimes (many-strata and
fixed-strata), their central limit theorems (Krewski--Rao and
triangular-array), and the conservative
with-replacement/finite-population correction (FPC) treatment are in Web
Appendix \S A (with the FPC calibration verified in Web Appendix \S F).

\textbf{Cross-fitted survey TMLE.} Our theorems cover the cross-fitted
Fully-Aware-CF estimator defined in Section 3.

\textbf{One-step expansion.} Let \(D(O;\eta_0,\Psi)\) denote the
per-unit influence function of Equation (\ref{eq:eif}), written \(D_i\)
there, evaluated at the true nuisances \(\eta_0=(Q_0,g_0)\) rather than
at the fitted \(Q_w^*\). The cross-fitted estimator admits the exact
decomposition \begin{equation*}
\widehat\psi_w-\Psi=\underbrace{(P_n^w-P_0)\,D(\cdot;\eta_0,\Psi)}_{S}\;+\;T_1\;+\;T_2,
\end{equation*} in which \(S\) is the survey-weighted average of the
fixed, true-nuisance influence function, \(T_1\) is an empirical-process
term in the fitted-minus-true nuisances, and \(T_2\) is a second-order
remainder. Four conditions, stated in full in Web Appendix \S A, control
these terms:

\begin{enumerate}
\item[(C1)] \emph{Neyman orthogonality and a product rate}: the influence function has zero derivative with respect to the nuisances at the truth, and $\lVert\widehat Q-Q_0\rVert_{L_2}\,\lVert\widehat g-g_0\rVert_{L_2}=o_p(m^{-1/2})$ (with each factor $o_p(m^{-1/4})$); then $T_2=o_p(m^{-1/2})$. Because the bounded per-PSU sample size forces $n=\Theta(m)$, a cube-root learner such as the highly adaptive lasso satisfies the per-factor clause and (C1) is arithmetically reachable even in the binding $m_h=2$ regime; what i.i.d.\ rate theory does not guarantee is that the within-PSU dependence preserves the learner's exponent (Web Appendix \S A, scale-bookkeeping remark).
\item[(C2)] \emph{PSU-level cross-fitting}: folds are formed of whole PSUs, assigned by a data-independent rule and balanced within every stratum (up to remainder PSUs whose total number across strata is $o(\sqrt m)$---automatic with $H$ fixed and ensured by cyclic allocation in the designs analyzed here; with $m_h=2$, leave-one-PSU-per-stratum), so the out-of-fold nuisances entering a held-out PSU's influence function---the outcome offset and the propensity---are independent of that PSU's fold given the training data. The single pooled fluctuation pair $(\epsilon_0,\epsilon_1)$ is fit on all folds and is \emph{not} out-of-fold; no independence is claimed for it: being two-dimensional, its contribution is controlled by a maximal inequality uniform over the fluctuation (Web Appendix \S A). Together with the factor rate in (C1), these give $T_1=o_p(m^{-1/2})$ \emph{without any Donsker} (entropy) condition on the Super Learner---requiring instead only the weaker finite-population \emph{no-memorization} condition (C4)---the central step of the argument.
\item[(C3)] \emph{Design nondegeneracy}: the design variance $\sigma_m^2$ of the leading term is nondegenerate on the PSU scale---$m\sigma_m^2$ converges to a positive limit. This is the only distributional limit assumed: the central limit theorem for the independent weighted cluster totals of the fixed influence function and the consistency of Equation (\ref{eq:var}) for $\sigma_m^2$ are then \emph{proved} in Web Appendix \S A (by Krewski--Rao arguments in the many-strata regime and a triangular-array argument in the fixed-strata regime), and, with the vanishing first-stage sampling fraction and bounded population cluster sizes rendering the model-layer deviation negligible (Web Appendix \S A), $S/\sigma_m\xrightarrow{d}\mathcal N(0,1)$.
\item[(C4)] \emph{Finite-population stability}: the realized out-of-fold nuisance does not \emph{memorize} the finite population---a no-memorization condition strictly weaker than a Donsker condition on the nuisance class, satisfied by fixed-entropy ensembles (parametric, finite Vapnik--Chervonenkis (VC), the highly adaptive lasso, or a cross-validated ensemble of these) or, for an arbitrary bounded learner, under the polynomially stronger sampling fraction $m^{3/2}/M_{\mathrm{pop}}\to0$ together with a mild first-stage regularity clause (Web Appendix \S A). This is the finite-population residue cross-fitting alone does not remove: cross-fitting makes a held-out fold independent of its out-of-fold nuisances, but the training PSUs remain in the realized population.
\end{enumerate}

\noindent Condition (C1) is an \emph{assumption} on the learners, not a
theorem: no oracle inequality is available for a survey-weighted,
clustered Super Learner, so we do not prove the product rate holds.
Fitting the out-of-fold nuisances \emph{unweighted} (Web Appendix \S B)
removes the survey weights from the learning step and so
\emph{motivates} the product rate via the ordinary oracle inequalities,
with two caveats that keep (C1) an assumption: the learner's rate must
transfer from the sampling to the population norm, which requires the
bounded relative weights (Web Appendix \S A, Proposition 1), and the
standard oracle inequalities are proved for i.i.d.~data, whereas even
the unweighted training observations remain dependent within PSUs. The
simulation therefore checks the realized rate along the complexity
ladder. Retaining the analogous term for the realized finite population
shows the same limit holds for \(\Psi_N\). Full statements and proofs
are given in Web Appendix \S A.

\textbf{Theorem 1 (Asymptotic normality).} Under conditions (C1)--(C4)
above (Web Appendix \S A), the cross-fitted survey TMLE satisfies
\((\widehat\psi_w-\Psi)/\widehat{\mathrm{se}}_w\xrightarrow{d}\mathcal N(0,1)\)
with \(\widehat{\mathrm{se}}_w\) the square root of Equation
(\ref{eq:var}); the Wald interval
\(\widehat\psi_w\pm z_{1-\alpha/2}\,\widehat{\mathrm{se}}_w\) thus
attains asymptotic coverage \(1-\alpha\) for the super-population ATE
\(\Psi\).

\textbf{Theorem 2 (Design-consistency of the variance).} Under the same
conditions, the estimator of Equation (\ref{eq:var}) evaluated on the
cross-fitted influence function is consistent for the design-based
variance \(\sigma_m^2\) that drives the limit of
\(\widehat\psi_w\)---the conditional design variance of the leading
\emph{true-nuisance} term, which like Theorem 1 presupposes the nuisance
consistency in (C1) and the finite-population stability condition (C4)
(Web Appendix \S A; the product rate is not used here); that is,
\(\widehat{\mathrm{se}}_w^2/\sigma_m^2\xrightarrow{p}1\).

In finite samples we refer the Wald statistic to a \(t\) distribution
with the survey degrees of freedom (the number of PSUs minus the number
of strata), the standard small-sample calibration used in the simulation
and application below.

\textbf{Scope: single-fit TMLE and the Donsker boundary.} The theorems
cover the cross-fitted estimator. A single-fit Super Learner TMLE---one
that fits the nuisances on the full sample and reuses them in the
influence function---is covered by our theory only in the parametric,
fixed-dimensional case (for example generalized-linear nuisances), where
a finite-dimensional maximal-inequality argument controls the
empirical-process term. Rich Super Learner libraries generally violate
even the Donsker condition, in which case \(T_1\) need not vanish and
the Wald interval can under-cover. The interpolating rung, moreover,
violates the product-rate condition (C1) by construction, so at L4
\emph{neither} the single-fit nor the cross-fitted estimator is covered
by Theorem 1; the cross-fitted estimator's at-or-above-nominal coverage
there is an empirical, design-specific observation---conservative in
direction in our experiments---not a guarantee. The single-fit
parametric case, the Donsker-regime behavior we expect but do not
establish, and the resulting practical recommendation are treated in
full in Web Appendix \S A (Remark on the single-fit estimator).

\textbf{Influence function, not efficient influence function.} Because
the independent units are clusters rather than individuals, the
individual-level function \(D\) is a valid influence function under the
known design but is not the efficient gradient of the clustered model;
the non-augmented weighted estimator is in general inefficient relative
to an augmented two-phase estimator \citep{robins1994estimation}. This
gap is not uniformly second order under clustering: the
influence-function design effect is a modest, NHANES-like 1.2--1.4 in
the headline designs, but reaches about 2.6 at the parametric rung and
about 3.4 at the most aggressive learner in a high-design-effect stress
design (Web Appendix \S F), so optimal two-phase augmentation could
deliver non-negligible precision gains where the clustered design effect
is large. We therefore use the term \emph{influence function}
throughout, rather than \emph{efficient influence function}, and leave
the optimal augmentation to future work.

\textbf{Calibrated weights.} For calibrated or post-stratified weights
such as the NHANES \texttt{WTMEC2YR}, the influence function acquires an
additional projection term; treating these weights as pure inverse
inclusion probabilities in Equation (\ref{eq:var}) is \emph{expected} to
be conservative in the usual calibration-linearization sense, in that it
does not credit the variance reduction afforded by calibration
\citep{saegusa2013weighted, breslow2007weighted}. We do not derive
variance consistency for calibrated weights: Theorems 1--2 are stated
for known inverse-probability weights under the idealized
with-replacement design above, so the NHANES application rests on that
approximation and the calibrated-weights-as-inverse-probability
convention.

\section{Simulation Study}\label{simulation-study}

We assess the estimators on synthetic data that reproduce the
stratification, clustering, and informative selection of a complex
survey, against a known population truth \citep{morris2019using}.

\begin{table}[t]\centering\small
\caption{Roadmap of the simulation study: each experiment, the question it addresses, and what it shows. The headline complexity ladder is Figure~\ref{fig:ladder} and Table~\ref{tab:sim}; full results for every experiment are in Web Appendix~\S D.}
\label{tab:simmap}
\begin{tabular}{@{}p{0.17\textwidth} p{0.63\textwidth} p{0.14\textwidth}@{}}
\toprule
Experiment & Question addressed, and what it shows & Location \\
\midrule
Headline complexity ladder & \emph{Are cross-fitting, the design-based variance, and the weights each necessary?} Single-fit coverage collapses at the interpolating rung while the cross-fitted estimator stays near nominal across the ladder; an i.i.d.\ variance and an unweighted estimate each under-cover. & Fig.~\ref{fig:ladder}, Table~\ref{tab:sim} \\
\addlinespace
Correctly-specified control & \emph{Is the lower-rung bias misspecification, not a pipeline defect?} Under a correctly specified data-generating process the cross-fitted estimator is unbiased with nominal coverage; the bias appears only under the Kang--Schafer design. & Web Table~S9 \\
\addlinespace
Cross-fitting vs.\ de-weighting; harmonized floor & \emph{Which ingredient restores coverage---cross-fitting, the weights, or the propensity floor?} Cross-fitting is the active ingredient: single-fit under-covers regardless of weighting or floor, and both cross-fitted arms recover. & Web Tables~S10, S12 \\
\addlinespace
Large-sample behavior & \emph{Is the single-fit collapse a finite-sample artifact that more data cures?} No---it is structural: single-fit coverage worsens as the number of clusters grows, while cross-fitting stays near nominal. & Web Table~S15 \\
\addlinespace
Product-rate diagnostic and $\sqrt{m}$ sweep & \emph{Does the assumed nuisance product rate hold empirically?} The $\sqrt{m}$-scaled product error is flat through the deployable rungs and inflates only at the interpolating extreme. & Web Tables~S13, S14 \\
\addlinespace
Type-I error and power & \emph{Does the primary estimator control type-I error at a true null?} Cross-fitting controls type-I with calibrated power; single-fit over-rejects at the interpolating rung and the design-naive arms over-reject throughout. & Web Table~S6 \\
\addlinespace
Double robustness & \emph{Is the estimator doubly robust in the two nuisances?} Point estimation retains the doubly robust structure---bias appears only when both nuisances are wrong---verified empirically; the formal theorems assume both nuisances consistent, and no design-based variance guarantee is claimed under single-nuisance misspecification. & Web Table~S11 \\
\addlinespace
External AIPW benchmark & \emph{Is the cross-fitting pattern specific to TMLE?} No---an independent survey-weighted AIPW exhibits the same single-fit collapse and cross-fit rescue. & Web Table~S7 \\
\addlinespace
Internal CV vs.\ cross-fitting on a realistic Super Learner library & \emph{Can internal cross-validation substitute for cross-fitting?} No---on a diverse Super Learner library built to standard practice (the deep forest just one of several learners, not a lone interpolator), only cross-fitting is calibrated; single-fit and internal cluster-cross-validation under-cover for both TMLE and AIPW. & Fig.~\ref{fig:ladder}(c,d); Web Table~S8 \\
\bottomrule
\end{tabular}
\end{table}

\subsection{Design}\label{design}

From a fixed finite population we take a two-stage stratified cluster
sample, selecting PSUs within strata and individuals within PSUs with
stratum-specific sampling fractions that make the design weights
genuinely unequal and the selection informative; the ATE on the
risk-difference scale is \(\Psi\approx0.21\). The headline coverage
study uses two designs (a high-design-effect stress variant, Design C,
is added in Web Appendix \S F): \emph{Design A} (\(H=10\) strata, six
PSUs per stratum, \(n=1{,}518\)), with many primary sampling units per
stratum, and \emph{Design B} (\(H=50\) strata, two PSUs per stratum,
\(n=2{,}034\)), which mirrors the many-strata, two-PSU structure of the
NHANES \texttt{SDMVSTRA}/\texttt{SDMVPSU} variables. The two designs sit
at finite design points of the two asymptotic regimes of Section 4:
Design A the fixed-strata regime (R2) and Design B the many-strata,
two-PSU regime (R1); in both limits the population PSU pool grows so the
first-stage sampling fraction vanishes, while at the frozen design
points it is fixed, where the no-correction variance of Equation
(\ref{eq:var}) is the calibrated, safe choice (Web Appendix \S A and
\S F). The resulting design effect on the influence function is modest
and NHANES-like (1.2--1.4). The full data-generating process---the
\(N\approx120{,}000\) super-population construction with its PSU-level
random intercepts, the Gauss--Hermite computation of the truth, the
exact inclusion fractions and variance components, and the
intracluster-correlation and design-effect audit---is in Web Appendix
\S C.

\subsection{The Super Learner complexity
ladder}\label{the-super-learner-complexity-ladder}

Our central device is a ladder of Super Learner libraries ordered by
\emph{function-class complexity} rather than by the number of learners.
A library can be enlarged indefinitely with smooth, low-complexity
learners without ever leaving the Donsker regime, so a count-based
ladder cannot expose the single-fit failure; only crossing into a
non-Donsker function class can. The four rungs L1--L4 run from a
generalized linear model to an aggressively grown, interpolating random
forest (Figure \ref{fig:ladder}; full library composition in Web
Appendix \S D), and the same library serves both the outcome and the
propensity model.

\subsection{Estimators compared}\label{estimators-compared}

Each sample is analyzed by five estimators that differ only in how they
use the design. \emph{Fully-Aware} (single-fit) uses survey-weighted
nuisances and targeting with the design-based variance of Equation
(\ref{eq:var}). \emph{Fully-Aware-CF} is the cross-fitted estimator
introduced earlier, identical except that the nuisances are fit
out-of-fold at the PSU level. \emph{Fully-Aware-CV} matches Fully-Aware
but lets the Super Learner choose its ensemble using cluster-aware
(PSU-respecting) cross-validation folds \citep{wieczorek2022k}; this
changes only the within-fit ensemble selection, not the cross-fitting
term, and is defined wherever the library has more than one learner (the
multi-learner rungs L2 and L3, and the diverse, standard-practice Super
Learner libraries L5--L7, which we call \emph{deployable}). Like
Fully-Aware, the CV arm is single-fit and not covered by Theorem 1 for
non-Donsker libraries; because the application configures this arm
differently (unweighted nuisances; Section 6), figures and tables label
both variants Fully-Aware-CV. \emph{Partially-Aware} keeps the
survey-weighted point estimate but uses an independence (i.i.d.)
variance, isolating the role of stratification and clustering.
\emph{Non-Aware} drops the weights throughout, isolating the role of
informative sampling. For each estimator, rung, and design we report,
over 1,000 replicates, the bias, the empirical standard deviation, the
mean estimated standard error, and the Wald coverage (referred to a
\(t\) distribution with the survey degrees of freedom), together with
the influence-function design effect.

\subsection{Results}\label{results}

\begin{figure}[th]
\includegraphics[width=1\linewidth,alt={a two-by-two grid of line plots; the top row shows coverage across L1--L4 for both designs, with Fully-Aware-CF staying high near 0.95 while single-fit Fully-Aware and Partially-Aware drop sharply to about 0.2 at L4 and Non-Aware under-covers throughout (lowest through L3); the bottom row shows coverage across the deployable libraries L5--L7 for both designs, with Fully-Aware-CF near 0.95 and the single-fit and internal-CV arms a few points below throughout.}]{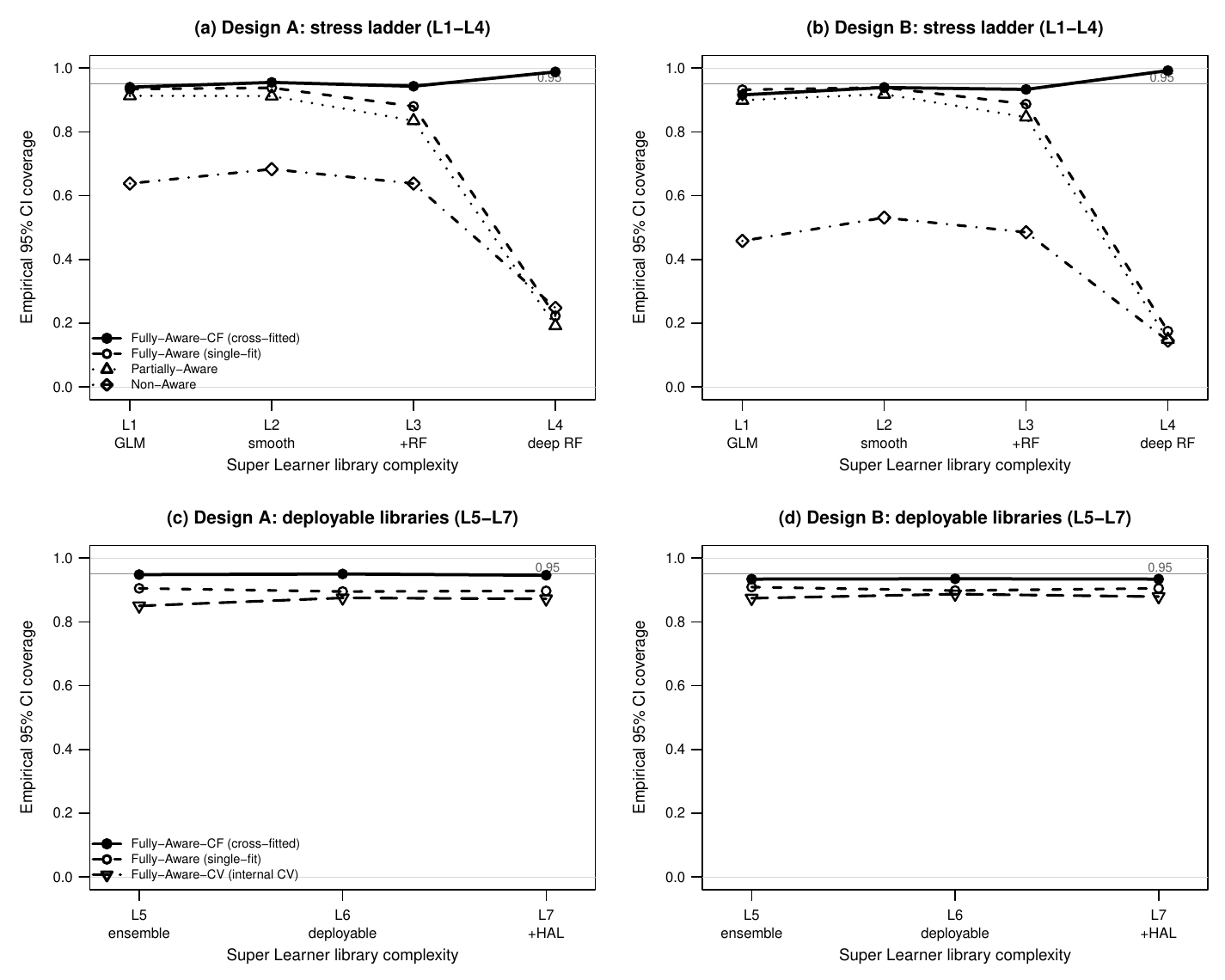} \caption{Empirical coverage of nominal 0.95 confidence intervals across the Super Learner library, by design (Design A, many PSUs per stratum; Design B, two PSUs per stratum, NHANES-like), over 1,000 replicates. Top row, panels (a) and (b): the single-learner stress ladder (L1 generalized linear model; L2 adds smooth learners; L3 adds a random forest; L4 a deep, interpolating random forest). The cross-fitted Fully-Aware-CF estimator stays close to nominal across all four rungs (about 0.91--0.99: at or near 0.94 at L2--L3, conservative at L4), whereas single-fit Fully-Aware and Partially-Aware track near 0.9 through L3 and then collapse to about 0.2 at L4; Non-Aware under-covers throughout, and worse in Design B. Bottom row, panels (c) and (d): three deployable Super Learner libraries built to standard practice -- diverse base learners with the deep forest just one member rather than a lone interpolator (L5, the three-learner ensemble GLM + MARS (multivariate adaptive regression splines) + deep random forest; L6, the five-learner library adding elastic net and gradient boosting; L7, L6 plus the highly adaptive lasso). Across all three deployable libraries and both designs only cross-fitting is calibrated (Fully-Aware-CF about 0.93--0.95), while single-fit Fully-Aware (about 0.89--0.91) and internal cluster-cross-validation Fully-Aware-CV (about 0.85--0.88) under-cover -- so even a diverse library in which cross-validation could down-weight the deep learner still needs cross-fitting. The internal-CV arm is shown only where the library has more than one learner. Lines are distinguished by type and plotting symbol for black-and-white reproduction.}\label{fig:ladder}
\end{figure}

Figure \ref{fig:ladder} shows coverage across the ladder for the four
estimators other than Fully-Aware-CV in both designs, and Table
\ref{tab:sim} reports bias, variability, and coverage at representative
rungs. Four findings emerge.

First, \emph{the single fit fails as the learners grow complex while the
cross-fit does not.} Single-fit Fully-Aware coverage is near nominal at
the smooth rungs (L1--L2), slips at L3, and collapses at the
interpolating L4 in both designs, driven by large bias and standard
errors far below the empirical spread (Table \ref{tab:sim}). The
cross-fitted estimator instead stays close to nominal across the entire
ladder---mildly below nominal at L1 in Design B, the familiar
finite-sample cost of sample splitting at a parametric rung, and mildly
conservative at L4, where its bias is negligible but its mean standard
error exceeds the empirical spread: in these replications an honest
widening from evaluating an under-smoothed learner out of fold rather
than a malfunction of the variance estimator, though the conservative
direction is a finite-sample observation, not a guarantee (Web Appendix
\S A). Because L4 violates the product rate (C1) by construction,
Theorem 1 covers neither arm at that rung; the cross-fitted arm errs
conservatively (intervals slightly too wide) while the single fit errs
anti-conservatively (intervals too narrow) in this design. More
carefully, the theory covers the cross-fitted estimator only where its
conditions (C1)--(C4) hold: the matched correctly-specified control (Web
Appendix \S D) is designed so that they plausibly do ((C1) itself
remains an assumption for data-adaptive learners; Section 4), and there
the cross-fitted arm is at or near nominal across L1--L3 (the single fit
additionally requires the Donsker condition, which already fails at L3).
The headline cells shown here use Kang--Schafer--misspecified working
nuisances, so consistency---and hence the product rate---is not
guaranteed even at L1--L3; the cross-fitted arm's near-nominal coverage
in these cells is an empirical observation rather than a direct
consequence of Theorem 1, and the small residual bias at the lower rungs
traces to that misspecification, not to the estimator---a matched
correctly-specified control reduces it to a negligible residual while
coverage stays nominal (Web Appendix \S D). The positivity signature
confirms the mechanism: at L4 the single-fit propensity estimates reach
the boundary \([0,1]\), whereas the cross-fitted propensities are held
off it by the imposed 0.05 floor.

Second, \emph{the design-based variance is necessary.} Partially-Aware
uses the correct weighted point estimate but an i.i.d. variance; it sits
roughly two to five points below Fully-Aware at L1--L3 and collapses
with it at L4, confirming that ignoring stratification and clustering
under-covers even when the point estimate is right.

Third, \emph{the weights are necessary.} Non-Aware under-covers
throughout (0.64--0.68 at L1--L3 in Design A and 0.46--0.53 at L1--L3 in
Design B, dropping to about 0.25 and 0.14 respectively at L4) and is
worse in the more informative Design B, because unweighted nuisances
estimate a sample rather than a population quantity.

Fourth, internal cross-validation does not substitute for cross-fitting:
at L3 the Fully-Aware-CV coverage (0.84 in Design A, 0.87 in Design B)
tracks the single-fit estimator rather than the cross-fitted one (0.94
and 0.93), so cluster-aware fold selection improves ensemble choice but
leaves the empirical-process term untouched; the same ordering holds
across three deployable Super Learner libraries built to standard
practice---diverse learners with the deep forest just one member, not a
lone interpolator (L5--L7; compositions in the caption and Web Appendix
\S D): only cross-fitting stays calibrated (about 0.93--0.95) while
single-fit Fully-Aware (about 0.89--0.91) and internal
cluster-cross-validation Fully-Aware-CV (about 0.85--0.88) under-cover
in both designs (Figure \ref{fig:ladder}c,d). The failure is thus not an
artifact of the pathological lone forest at L4---even a properly
specified, diverse library needs cross-fitting---and it also holds for
the unweighted internal-CV recipe the application uses (Web Appendix
\S D).

\begin{table}[t]
\centering
\caption{Simulation performance across the Super Learner complexity ladder (L1--L4; 1,000 replicates): bias, empirical standard deviation (SD), mean estimated standard error (SE), the ratio of the mean SE to the empirical SD (SE/SD; near 1 is calibrated, below 1 anticonservative, above 1 conservative), and 95\% coverage with its Monte Carlo standard error in parentheses, for Design A (many PSUs per stratum) and Design B (two PSUs per stratum, NHANES-like). Fully-Aware-CV, defined wherever the library has more than one learner, is shown here at the multi-learner rungs L2 and L3; it stays near nominal at the still-Donsker L2 and diverges from the cross-fitted estimator at the non-Donsker L3. Monte Carlo standard errors are at most $0.005$ for bias and $0.004$ for the empirical standard deviation across all cells, so estimator differences below about $0.01$ at a given rung are within simulation noise. The data-generating process is the Kang--Schafer misspecified design; the correctly-specified control and complete results are in Web Appendix \S D.}
\label{tab:sim}
\resizebox{\ifdim\width>\linewidth\linewidth\else\width\fi}{!}{
\begin{tabular}{ll rrrrr rrrrr}
\toprule
\multicolumn{12}{@{}l}{\footnotesize Kang--Schafer misspecified DGP (\texttt{model\_type=complex}); correctly-specified control in Web Appendix~D} \\
\addlinespace[2pt]
 & & \multicolumn{5}{c}{Design A} & \multicolumn{5}{c}{Design B} \\
\cmidrule(lr){3-7}\cmidrule(lr){8-12}
Library & Estimator & Bias & SD & SE & SE/SD & Cov & Bias & SD & SE & SE/SD & Cov \\
\midrule
L1 (GLM) & Fully-Aware (single-fit) & +0.018 & 0.043 & 0.044 & 1.02 & 0.934\,{\scriptsize(0.008)} & +0.018 & 0.034 & 0.034 & 1.00 & 0.932\,{\scriptsize(0.008)} \\
 & Fully-Aware-CF & +0.019 & 0.044 & 0.047 & 1.05 & 0.940\,{\scriptsize(0.007)} & +0.020 & 0.036 & 0.036 & 1.00 & 0.916\,{\scriptsize(0.009)} \\
 & Partially-Aware & +0.018 & 0.043 & 0.040 & 0.92 & 0.913\,{\scriptsize(0.009)} & +0.018 & 0.034 & 0.032 & 0.93 & 0.899\,{\scriptsize(0.009)} \\
 & Non-Aware & +0.031 & 0.039 & 0.024 & 0.62 & 0.638\,{\scriptsize(0.015)} & +0.044 & 0.028 & 0.021 & 0.76 & 0.458\,{\scriptsize(0.016)} \\
\addlinespace
L2 (smooth) & Fully-Aware (single-fit) & +0.010 & 0.043 & 0.043 & 0.99 & 0.938\,{\scriptsize(0.008)} & +0.010 & 0.034 & 0.034 & 0.98 & 0.939\,{\scriptsize(0.008)} \\
 & Fully-Aware-CF & +0.013 & 0.044 & 0.048 & 1.07 & 0.955\,{\scriptsize(0.007)} & +0.015 & 0.036 & 0.036 & 1.01 & 0.939\,{\scriptsize(0.008)} \\
 & Fully-Aware-CV & +0.008 & 0.044 & 0.045 & 1.01 & 0.938\,{\scriptsize(0.008)} & +0.009 & 0.034 & 0.034 & 0.98 & 0.943\,{\scriptsize(0.007)} \\
 & Partially-Aware & +0.010 & 0.043 & 0.039 & 0.89 & 0.912\,{\scriptsize(0.009)} & +0.010 & 0.034 & 0.031 & 0.90 & 0.918\,{\scriptsize(0.009)} \\
 & Non-Aware & +0.025 & 0.039 & 0.024 & 0.62 & 0.683\,{\scriptsize(0.015)} & +0.038 & 0.028 & 0.021 & 0.74 & 0.531\,{\scriptsize(0.016)} \\
\addlinespace
L3 (+RF) & Fully-Aware (single-fit) & +0.008 & 0.043 & 0.036 & 0.84 & 0.880\,{\scriptsize(0.010)} & +0.009 & 0.034 & 0.028 & 0.84 & 0.887\,{\scriptsize(0.010)} \\
 & Fully-Aware-CF & +0.012 & 0.044 & 0.047 & 1.06 & 0.943\,{\scriptsize(0.007)} & +0.015 & 0.036 & 0.036 & 1.00 & 0.933\,{\scriptsize(0.008)} \\
 & Fully-Aware-CV & +0.009 & 0.047 & 0.034 & 0.73 & 0.840\,{\scriptsize(0.012)} & +0.009 & 0.035 & 0.028 & 0.80 & 0.870\,{\scriptsize(0.011)} \\
 & Partially-Aware & +0.008 & 0.043 & 0.032 & 0.74 & 0.835\,{\scriptsize(0.012)} & +0.009 & 0.034 & 0.026 & 0.77 & 0.846\,{\scriptsize(0.011)} \\
 & Non-Aware & +0.024 & 0.039 & 0.022 & 0.55 & 0.638\,{\scriptsize(0.015)} & +0.037 & 0.028 & 0.019 & 0.67 & 0.485\,{\scriptsize(0.016)} \\
\addlinespace
L4 (deep RF) & Fully-Aware (single-fit) & +0.089 & 0.118 & 0.023 & 0.20 & 0.223\,{\scriptsize(0.013)} & +0.098 & 0.107 & 0.017 & 0.16 & 0.175\,{\scriptsize(0.012)} \\
 & Fully-Aware-CF & +0.002 & 0.046 & 0.066 & 1.45 & 0.988\,{\scriptsize(0.003)} & +0.006 & 0.036 & 0.052 & 1.44 & 0.992\,{\scriptsize(0.003)} \\
 & Partially-Aware & +0.089 & 0.118 & 0.020 & 0.17 & 0.192\,{\scriptsize(0.012)} & +0.098 & 0.107 & 0.016 & 0.15 & 0.149\,{\scriptsize(0.011)} \\
 & Non-Aware & +0.046 & 0.038 & 0.012 & 0.32 & 0.248\,{\scriptsize(0.014)} & +0.052 & 0.030 & 0.011 & 0.35 & 0.145\,{\scriptsize(0.011)} \\
\bottomrule
\end{tabular}
}
\end{table}

\section{Application: Four Illustrations in
NHANES}\label{application-four-illustrations-in-nhanes}

We illustrate the estimators on four contemporary questions in the
National Health and Nutrition Examination Survey (NHANES, 2007--2018), a
complex multistage probability survey of the civilian,
non-institutionalized United States population. The four were chosen to
span the range of exposure prevalence and overlap, and to show that
respecting the design can change what the analyst would
report---revealing an understated association, dissolving an apparent
one, or stabilizing a rare-exposure estimate: (E1) short sleep (\(<7\) h
vs 7--9 h) and obesity; (E2) household food insecurity and depression
(PHQ-9-defined); (E3) e-cigarette use (ever vs never) and hypertension;
and (E4) a history of gestational diabetes (GDM) and later-life
hypertension. All four domains are adults aged 20 and over (E4 further
restricts to ever-pregnant women without prevalent diabetes); precise
eligibility per example is in Web Appendix \S E. The four analyses are
methodological illustrations, not definitive epidemiologic studies (Web
Appendix \S E discusses the status of each adjustment set). Each
estimand is the population ATE on the risk-difference scale, and each is
estimated with the full set of arms defined in Section 5 (Non-Aware,
Partially-Aware, Fully-Aware, Fully-Aware-CV, and the primary
Fully-Aware-CF). The CV arm here uses unweighted nuisances (like
Fully-Aware-CF), differing only in internal cross-validation versus
PSU-level cross-fitting, and is not covered (Web Appendix \S F).

\textbf{Data and design.} Each analysis is a \emph{sub-population}
(domain) of the full NHANES sample: we build the survey design (pooled
MEC weights, masked strata, and primary sampling units) on the full
sample and then subset it, rather than deleting records first, so the
full stratum and PSU structure is preserved and partially intersecting
PSUs contribute correctly to the domain variance. The weight pooling,
\texttt{svydesign} construction, the resulting survey degrees of
freedom, the hypertension definition and its cross-cycle measurement
caveat, and the imputation mechanics are detailed in Web Appendix \S E.

\textbf{Covariates.} For each question we adjust only for pre-exposure
common causes and exclude descendants of the exposure or outcome; the
same variable can play different roles across questions (diabetes is a
confounder in E3 but a post-exposure variable in E4), as the per-example
adjustment sets encode. Full variable definitions, adjustment sets, the
unmeasured-confounder proxies, and the over-adjustment and proxy
sensitivity analyses are in Web Appendix \S E and \S F.

\begin{figure}[th]
\includegraphics[width=1\linewidth,alt={four forest plots, one per question; in every panel the design-aware markers lie at similar horizontal positions while the cross-fitted interval is comparable to or narrower than the other design-aware arms; in the e-cigarette panel only the Non-Aware interval sits entirely to the right of zero, whereas in the short-sleep and food-insecurity panels all five intervals exclude zero.}]{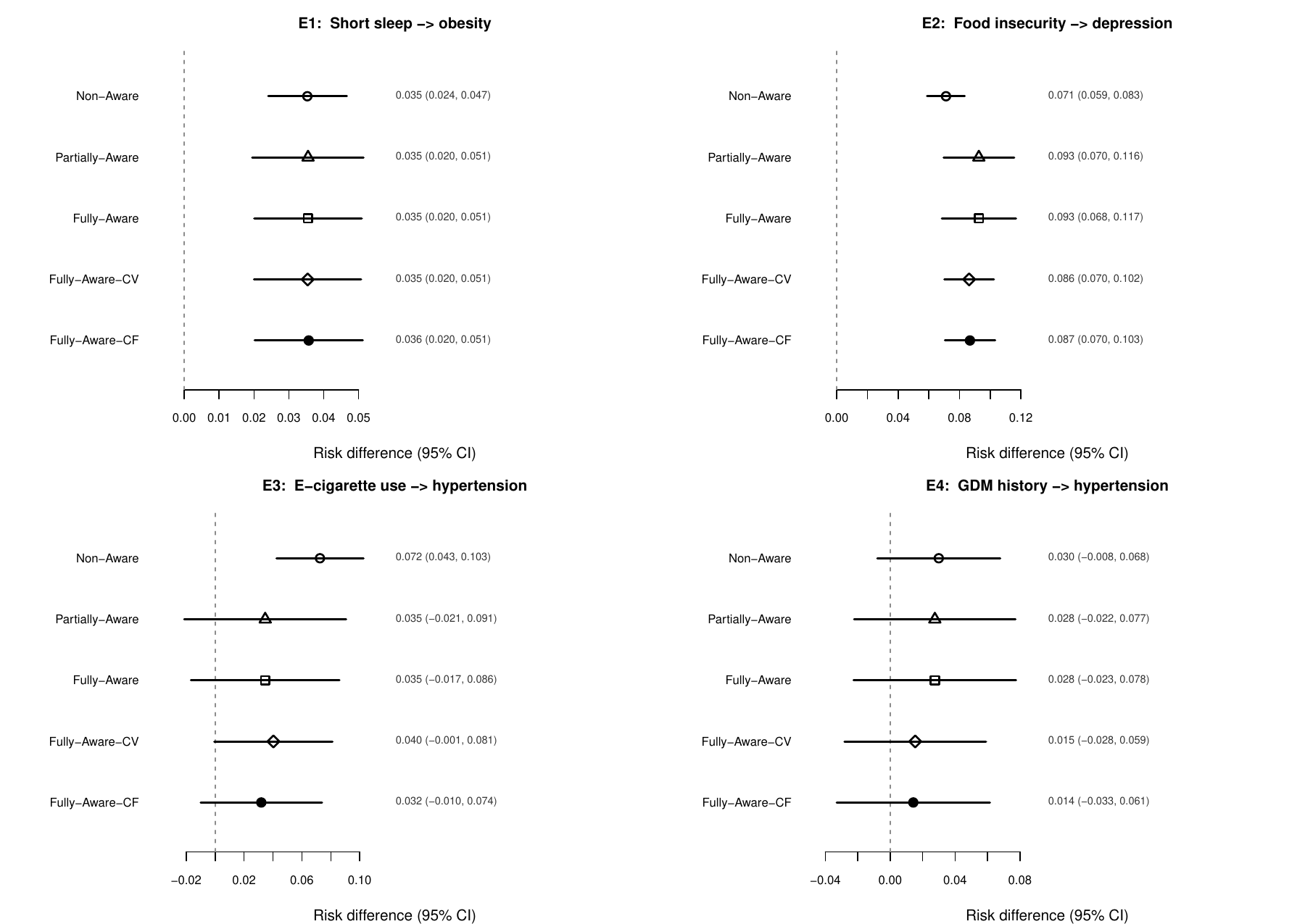} \caption{Population average treatment effect (risk difference, with 95\% confidence intervals---design-based for the survey-aware arms, naive for Non-Aware) for four NHANES illustrations (E1--E4), each estimated by five TMLE variants spanning the design-awareness ladder. Within each panel the point estimates are similar across the design-aware arms, though survey-weighting shifts the estimate away from the Non-Aware one in E2 and E3; the design-aware intervals are wider than the naive (Non-Aware) one, though the primary cross-fitted interval is no wider (at three-decimal precision) than the single-fit design-aware arms (the arms are ordered Non-Aware to the recommended primary Fully-Aware-CF, which is plotted with a filled marker; the plotted points are the multiple-imputation estimates ($m=40$), identical to Table~\ref{tab:nhanes}), and in E3 the Non-Aware interval excludes zero while the design-aware intervals do not.}\label{fig:nhanes}
\end{figure}

\textbf{Results.} Figure \ref{fig:nhanes} and Table \ref{tab:nhanes}
report the four analyses; both display the same multiple-imputation
values (\(m=40\))---the figure as a forest plot for visual comparison
across the design-awareness ladder, the table as the auditable
record---and a matched single-imputation cross-check (Web Appendix \S F)
confirms that the primary-arm conclusions are stable to the imputation
count in E1, E2, and E4; only E3 sits on the significance boundary,
where the primary interval covers zero under the headline multiple
imputation and marginally excludes it under single imputation (Web
Appendix \S F). The point estimates are broadly stable across arms; in
these four analyses the primary cross-fitted interval is, at the
reported three-decimal precision, no wider than the single-fit
design-aware arms in any example and narrower than both in E2, E3, and
E4 (the CV arm, sharing its unweighted nuisances, is comparable
throughout), though the simulation shows the cross-fitted estimator can
be mildly conservative (Section 5), so this width comparison is specific
to these designs rather than a general precision guarantee. What the
analyst would report nonetheless depends on how much of the design is
respected. In E1, where the exposure is common and overlap is excellent,
all arms agree and the design-aware interval is roughly a third wider
than the naive one. In E2, ignoring the weights \emph{understates} the
association, though both intervals exclude zero, so the weights change
the magnitude rather than the qualitative finding. In E3 the opposite
occurs: the Non-Aware analysis returns an apparently significant
association, but survey-weighting moves the estimate to an interval
covering zero and every design-aware arm agrees---the apparent
association is removed at the weighting step, not by cross-fitting. In
E4 the exposure is rare and overlap is poorest: the weighted exposed
mass below the imposed 0.05 propensity floor is about 21\%, against only
3--4\% in E2 and E3 (Web Table S29)---so there the primary estimate is
read as the overlap-restricted ATE (Web Appendix \S F). The single-fit
propensities reach 0.003 while the cross-fitted ones are held at or
above the floor, and the primary risk difference is near zero. We report
Fully-Aware-CF as our recommended estimator because it is the estimator
the theory of Section 4 covers for machine-learning nuisances; the full
five-arm comparison, a generalized-linear-tier calibration of the design
variance, and a deep-learner ladder are in Web Appendix \S F. That
ladder confirms the simulation's single-fit failure on real data too:
re-estimated up to a deep interpolating random forest, the single-fit
estimator breaks down in every example---inflated, swung, and even
sign-flipped estimates (Web Appendix \S F)---while the cross-fitted
estimator stays stable. Because the truth is unknown on a single sample,
this corroborates the point-estimate instability and positivity
mechanism rather than coverage (which the Section 5 simulation
confirms). A roadmap of the robustness and sensitivity analyses is in
Table \ref{tab:webF_sensmap}.

\begin{table}[t]
\centering
\caption{Four NHANES illustrations (2007--2018): for each, the analytic sample size and unweighted exposure prevalence, the population ATE (risk difference, 95\% CI) under all five design-awareness arms---Non-Aware, Partially-Aware, Fully-Aware (single-fit), Fully-Aware-CV, and the primary Fully-Aware-CF (bold)---pooled over $m=40$ multiple imputations of the missing covariates by Rubin's rules (matched single-imputation estimates in Web Appendix \S F), and the minimum estimated propensity score for the single-fit versus cross-fitted estimators (a positivity check). The point estimate is broadly stable across arms, and where the substantive conclusion shifts (E3) it shifts at the Non-Aware $\to$ Partially-Aware (survey-weighting) step, not at cross-fitting. In the primary arm the out-of-fold propensities are truncated at the 0.05 floor, which the single-fit estimates undershoot where overlap is poor (E2, E3, E4); in E4, where the floor binds materially, the primary estimate targets the overlap-restricted ATE (Web Table~S32). In this application Fully-Aware-CV uses unweighted nuisances (matching Fully-Aware-CF); the simulation's Fully-Aware-CV instead keeps survey-weighted nuisances (Web Appendix \S F).}
\label{tab:nhanes}
\resizebox{\ifdim\width>\linewidth\linewidth\else\width\fi}{!}{
\begin{tabular}{@{}l cccc@{}}
\toprule
Estimator & E1 & E2 & E3 & E4 \\
& sleep $\to$ obesity & food insec.\ $\to$ depr. & e-cig $\to$ HTN & GDM $\to$ HTN \\
$n$ (exposed) & 30{,}983 (37\%) & 29{,}359 (20\%) & 10{,}736 (18\%) & 10{,}956 (6\%) \\
\midrule
\multicolumn{5}{@{}l}{\emph{Risk difference (95\% CI), multiple imputation $m{=}40$}}\\
Non-Aware & 0.035 (0.024, 0.047) & 0.071 (0.059, 0.083) & 0.072 (0.043, 0.103) & 0.030 ($-$0.008, 0.068) \\
Partially-Aware & 0.035 (0.020, 0.051) & 0.093 (0.070, 0.116) & 0.035 ($-$0.021, 0.091) & 0.028 ($-$0.022, 0.077) \\
Fully-Aware & 0.035 (0.020, 0.051) & 0.093 (0.068, 0.117) & 0.035 ($-$0.017, 0.086) & 0.028 ($-$0.023, 0.078) \\
Fully-Aware-CV & 0.035 (0.020, 0.051) & 0.086 (0.070, 0.102) & 0.040 ($-$0.001, 0.081) & 0.015 ($-$0.028, 0.059) \\
\textbf{Fully-Aware-CF} & 0.036 (0.020, 0.051) & 0.087 (0.070, 0.103) & 0.032 ($-$0.010, 0.074) & 0.014 ($-$0.033, 0.061) \\
\midrule
min $\hat g$ (single / CF) & 0.13 / 0.13 & 0.004 / 0.05 & 0.002 / 0.05 & 0.003 / 0.05 \\
\bottomrule
\end{tabular}
}
\end{table}

\begin{table}[t]\centering\small
\caption{Roadmap of the robustness and sensitivity analyses (Web Appendices F and G). The first group stresses the method on synthetic data with known truth; the second checks that the four NHANES conclusions are robust.}
\label{tab:webF_sensmap}
\begin{tabular}{@{}p{0.21\textwidth} p{0.59\textwidth} p{0.13\textwidth}@{}}
\toprule
Analysis & Question addressed, and what it shows & Location \\
\midrule
\multicolumn{3}{@{}l}{\emph{Method robustness (simulation, known truth)}}\\
\addlinespace[2pt]
Informative selection & Do de-weighted out-of-fold fits hold when selection is informative beyond the confounders? In this DGP, empirically yes---coverage holds and bias declines across the sweep, matching weighted out-of-fold fits; a stress test outside the $S\perp(A,Y)\mid C$ theorem, not a general guarantee. & Web Table~S36 \\
\addlinespace
Finite-population correction & Does omitting the FPC over-cover when the sampling fraction is non-negligible? No---omitting it is calibrated and coverage is flat in the fraction. & Web Table~S25 \\
\addlinespace
High design-effect stress & Does the clustering-aware SE hold when the design effect is large? Yes---the primary estimator stays near nominal (conservative at the interpolating rung) while the design-naive arms collapse. & Web Table~S29 \\
\addlinespace
Jackknife vs.\ linearization & Is the inexpensive linearization SE optimistic relative to a resampling estimator? No---the delete-one-PSU jackknife and the linearization SE agree. & Web Table~S30 \\
\addlinespace
\multicolumn{3}{@{}l}{\emph{NHANES real-data sensitivity}}\\
\addlinespace[2pt]
Multiple imputation ($m=40$) & Are the conclusions an imputation-count artifact? Largely---multiple imputation and single imputation agree in E1, E2, and E4; in E3 the primary Fully-Aware-CF (and non-primary CV) interval sits on the significance boundary, covering zero under the headline multiple imputation and marginally excluding it under single imputation. & Web Table~S24 \\
\addlinespace
Real-data cross-fit isolation & Is the real-data correction driven by cross-fitting rather than the floor? The floor leaves the single-fit estimate unchanged; cross-fitting drives the correction. & Web Table~S35 \\
\addlinespace
Covariate-set and design check & Do the conclusions survive alternative adjustment sets and the pooled-strata assumption? Largely---E1, E2, and E4 are stable and the pooled-strata assumption holds; the E3 null is robust to dropping diabetes but turns borderline-significant when BMI is dropped (a strong outcome predictor whose adjustment shifts this borderline estimate), so the fully-adjusted design-aware estimate covering zero remains primary. & Web Table~S37 \\
\addlinespace
Propensity-floor sensitivity & Does loosening the $0.05$ floor move the rare-exposure estimate? No---it moves negligibly and its interval covers zero. & Web Table~S33 \\
\addlinespace
Share at the floor & Where does the floor bind (the overlap-restricted estimand)? Materially only for the rare exposure. & Web Table~S32 \\
\addlinespace
Propensity overlap & Where does single-fit positivity fail on real data? Only at the interpolating rung does the single-fit propensity reach the boundary; cross-fitting stays bounded. & Web Table~S34 \\
\addlinespace
Variance-method benchmark & Do cheaper design-based standard errors match a full survey bootstrap on real data, and at what cost? Yes---the linearization, jackknife, balanced repeated replication (BRR) and Fay's BRR agree with the $B{=}500$ survey bootstrap at a small fraction of its cost. & Web Tables~S38--S39 \\
\bottomrule
\end{tabular}
\end{table}

\section{Discussion}\label{discussion}

\textbf{Contributions and application}: Building on prior work adapting
causal inference to complex surveys \citep{nattino2025causal}, we give a
survey-weighted, cross-fitted TMLE for the population ATE that is
asymptotically normal with a design-consistent linearization variance
under the conditions of Section 4---centrally the assumed nuisance
product rate---and show, in theory and in simulations spanning a
many-PSU and an NHANES-like design, that with flexible machine-learning
nuisances single-fit survey TMLE can lose valid variance and
cross-fitting at the primary-sampling-unit level restores it.
Cross-fitting (Fully-Aware-CF) removes the Donsker requirement that rich
libraries violate---at the cost only of a weaker finite-population
no-memorization condition (Web Appendix \S A)---but not the rate
assumption. This closes the gap that has made survey-causal
\emph{variance} estimation---as opposed to the well-studied handling of
weights for point estimation---a persistent challenge often met only by
computationally intensive bootstraps \citep{austin2018propensity}: the
linearization standard error is formal, cheap, and integrated with
complex-survey theory, so confidence intervals and tests, not just point
estimates, are trustworthy. Across four contemporary NHANES questions
(Section 6) respecting the design altered what the analyst would
report---most sharply in e-cigarette use and hypertension, where a naive
analysis's apparently significant association was removed by
survey-weighting (the only conclusion that changes, and at the weighting
step, not at cross-fitting), and elsewhere through a larger weighted
food-insecurity--depression association, a cross-fitting-bounded
positivity near-violation under a rare exposure, and materially wider
but unchanged intervals where overlap was excellent. In each case the
primary cross-fitted estimate is the one our theory supports, and naive
analyses can overstate precision and risk erroneous conclusions
\citep{korn1999analysis}. The theory and implementation extend readily
to other complex survey datasets.

\textbf{Practical implications}: We recommend cross-fitted survey TMLE
with the design-based variance as the default when machine-learning
nuisances are used, using PSU-level folds (leave-one-PSU-per-stratum
when only two PSUs are sampled; the recipe is in Web Appendix \S B).
This is not an artifact of TMLE: on a deployable multi-learner library,
an independent survey-weighted augmented inverse-probability-weighted
estimator showed the same single-fit and internal-cross-validation
under-coverage that only cross-fitting removed, so the pattern reflects
the estimation problem rather than our choice of estimator (Web Appendix
\S D) \citep{zheng2011cross, chernozhukov2018double}.

\textbf{Limitations and future work}: The first limitation is intrinsic
to the theory: the product-rate condition (C1) is an assumption on the
learners, not something the framework can verify. It is plausible when
the covariates are bounded, the weight variation is modest, and the
learners are regularized---settings in which the ordinary (unweighted,
i.i.d.) oracle inequalities that motivate the rate are most
credible---and it fails by construction for interpolating learners such
as the L4 forest (minimum node size one); no oracle inequality currently
exists for survey-weighted or within-PSU-dependent training data, so the
diagnostic of Web Appendix \S D probes the realized rate but cannot
confirm it. A second theory-scope caveat concerns the weights: the
formal results assume a known-probability design, whereas NHANES weights
are calibrated and we treat them as inverse-probability
weights---conservatively, though not provably so for the cross-fitted
estimator (Section 4). Beyond this, our simulations rely on specific
data-generating mechanisms, and relative performance may differ under
other misspecifications or designs. Our four NHANES illustrations,
though spanning common and rare exposures, come from a single survey
program in one country, so generalization to other complex surveys
remains to be demonstrated. As in any observational study, residual
confounding from unmeasured or imperfectly measured factors remains a
concern, and the cross-sectional design limits temporal ordering and
time-varying confounding.

\textbf{Conclusion}: Valid population-level causal inference from
complex survey data requires a design-aware approach that accounts for
weights, stratification, and clustering---particularly in variance
estimation---and we show that cross-fitted survey TMLE delivers it using
off-the-shelf software. This closes a key gap between modern
doubly-robust causal inference and the complexities of real-world,
population-based health data.

\backmatter

\section*{Acknowledgements}

This research was supported in part through computational resources from
Advanced Research Computing at the University of British Columbia. The
author declares no conflicts of interest. During this work, the author
used AI-based tools (large language models) to assist with the analysis
and simulation code, text editing, and checking derivations; the author
verified all outputs and takes full responsibility for the
content.\vspace*{-8pt}

\section*{Data Availability}

The data are publicly available from the U.S. National Center for Health
Statistics (NHANES, \url{https://www.cdc.gov/nchs/nhanes/}). Code for
the method, simulation, and NHANES analysis is available at
\url{https://github.com/ehsanx/Survey-TMLE-codes}, with an interactive
companion at \url{https://ehsanx.github.io/survey-tmle-app/}. A
convenience R package, \texttt{svytmle}, is available at
\url{https://github.com/ehsanx/svytmle}.\vspace*{-8pt}

\section*{Supplementary Material}

Web Appendices A--G, Web Tables S1--S39, and Web Figures S1--S5 are
provided in the Supplementary Material accompanying this
preprint.\vspace*{-8pt}

\bibliographystyle{abbrvnat}
\bibliography{ref}

\label{lastpage}

\end{document}